\documentclass[10pt, conference, compsocconf]{IEEEtran}
\ifCLASSINFOpdf
\else
\fi

\usepackage{dsfont}
\usepackage{subfig}
\usepackage{rotating}
\usepackage[utf8]{inputenc}
\usepackage{array}
\usepackage{float}
\usepackage{placeins}

\begin{document}
%
\title{Implementing Noise with Hash functions for Graphics Processing Units}


\author{\IEEEauthorblockN{Matias Valdenegro-Toro}
\IEEEauthorblockA{Departamento de Informatica\\
Universidad Tecnologica Metropolitana\\
Santiago, Chile\\
matias.valdenegro@gmail.com}
\and
\IEEEauthorblockN{Hector Pincheira Conejeros}
\IEEEauthorblockA{Departamento de Informatica\\
Universidad Tecnologica Metropolitana\\
Santiago, Chile\\
hpinche51@gmail.com}
}


%


\maketitle

\begin{abstract}
We propose a modification to Perlin noise which use computable hash functions instead of textures as lookup tables.
We implemented the FNV1, Jenkins and Murmur hashes on Shader Model 4.0 Graphics Processing Units for noise generation.
Modified versions of the FNV1 and Jenkins hashes provide very close performance compared to a texture based Perlin noise implementation. Our noise modification enables noise function evaluation without any texture fetches, trading computational power for memory bandwidth.
\end{abstract}


%
\IEEEpeerreviewmaketitle

\section{Introduction}

Noise is a primitive function used in computer graphics to create real-looking procedural content and textures. It was introduced by Perlin \cite{imageSynthesizer} and it is the standard implementation for noise. The noise function returns a pseudorandom deterministic scalar output based on its n-dimensional input.

A noise function has some desirable features \cite{glslSpec}, such as:

\begin{enumerate}
 \item Continuous in its domain.
 \item A defined output domain, usually [-1, 1].
 \item An average of zero.
 \item Statistically invariant to transformations on its domain.
 \item Band limited in frequency.
\end{enumerate}

The original Perlin Noise algorithm was suited to a CPU implementation, and uses two lookup tables. A permutation table is used as a hashing function and a gradient table. Accessing them in Graphics Processing Unit (GPU) or massive parallel architectures can be a bottleneck, as the noise function can be used several times per processed fragment.

Removing the dependency on lookup tables is a difficult matter, as they provide the necessary entropy to generate a pseudorandom output. A pure computable noise function would be valuable for a hardware and/or a GPU implementation.

This paper proposes modifications to Perlin Noise which makes it purely computable, replacing both lookup tables with functions computed at runtime. This enables a fast GPU implementation, using the OpenGL Shading Language (GLSL).\\


\section{Previous Work}

On \cite{improvedPerlin}, Perlin introduced modifications to its classic Perlin Noise, using higher order interpolants remove discontinuities in the second derivate, which produced artifacts, and a new gradient distribution which hides some lattice-aligned artifacts.

There are several GPU implementations of Perlin Noise, such as baking a 1D/2D/3D noise texture and sampling to get noise values \cite{olanoModifiedNoise}, or implementing the complete algorithm, using textures to store the lookup tables \cite{implementingImprovedPerlin}.

Olano \cite{olanoModifiedNoise} proposed a modification to Perlin Noise using a LCG-based hash function, and an alternate gradient distribution, which produced noise with a period of 61 units. His proposal was aimed to a GPU Shader Assembly implementation.

\section{Generalized Perlin Noise}

Perlin noise uses a function $\mathds{N}^n \rightarrow \mathds{R}^n$ to assign every point in a integer lattice space a gradient vector of the same dimension. Perlin calculates this gradient using a precalculated gradient table ($\mathds{N} \rightarrow \mathds{R}^n$), indexed with the aid of a precalculated permutation table ($\mathds{N} \rightarrow \mathds{N}$):

\begin{verbatim}
int permute(int x, int y, int z)
{
    int px = permTable[x];
    int py = permTable[y + px];

    return permTable[py + z];
}

vec3 gradient(int x, int y, int z)
{
    return gradTable[permute(x, y, z)];
}
\end{verbatim}

There is no dependency between this gradient generation methodology and the Perlin noise algorithm \cite{imageSynthesizer}. Any other random generation method should be enough.

\section{Modern Graphics Processing Units}

GPUs have evolved from a fixed function programming model to a programmable model, in which the developer can execute code in defined stages of the graphics pipeline to achieve different effects.

GPU features are defined by shader model versions, defined between Microsoft and Hardware Vendors. The most recent version as this writing is Shader Model 4.0, which supports several features \cite{direct3D10System} useful for this paper:

\begin{enumerate}
 \item Full support for signed and unsigned integers, and bit operations on them.
 \item Unfiltered texture fetches.
 \item Texture fetches with pixel coordinates.
 \item Unlimited number of executed instructions.
\end{enumerate}

Integer and bit operation support is required to implement any common hashing function. This functionality is accessed through the OpenGL Shading Language.

The GL\_EXT\_gpu\_shader4 \cite{extGpuShader4} OpenGL extension exposes this funcionality for Shader Model 4.0 hardware. This extension has been integrated into the core OpenGL specification in version 3.0 \cite{glSpec}.

\section{Proposed changes to Perlin Noise}

We propose changing Perlin's gradient generation with a real hashing function, evaluated at runtime without lookup tables.

Using a hash function $\mathds{N} \rightarrow \mathds{N}$, it is evaluated on each component of the noise function input, but linked to the previous component evaluation in a similar way Perlin linked to its permutation evaluation. Then a n-dimensional integer vector is constructed, and used to evaluate a trigonometric function, converting the integer vector into a floatin point vector, finally yielding the n-dimensional gradient.

\begin{verbatim}
vec2 gradient(ivec2 p)
{
    int x = hash(p.x);
    int y = hash(x + p.y);

    return sin(vec2(x + y, y + y));
}

vec3 gradient(ivec3 p)
{
    int x = hash(p.x);
    int y = hash(x + p.y);
    int z = hash(y + p.z);

    return sin(vec3(z + x, z + y, z + z));
}
\end{verbatim}

Later the gradient is used normally with the Perlin Noise algorithm. For the hash function we chose 3 candidates, the Fowler-Noll-Vo-1 (FNV1), Murmur and Jenkins hashes. Criteria for the hash function selection is:\\

\begin{itemize}
 \item Small code footprint.
 \item Small execution time.
 \item Not a cryptographycally secure hash (due to execution time constraints).
\end{itemize}

\subsection{The FNV1 hash}

FNV is a hash function created by Fowler, Noll, and Vo \cite{fnv}. The hash is defined for power of two output bitsizes, starting from 32 bits to 1024 bits. It uses two magic numbers, the FNV offset basis and the FNV prime, both dependant on the output size. The pseudocode for the hash follows:

\begin{verbatim}
int hash(int input)
{
    int ret = fnvOffsetBasis;

    for each byte i in input {
        ret = ret * fnvPrime;
        ret = ret ^ i;
    }

    return ret;
}
\end{verbatim}

\subsection{The Murmur Hash}

Murmur is a hash function created by Appleby \cite{murmur}, which claims to have a excellent distribution, excellent avalanche and excellent collision resistance. It processes 32-bit blocks and has output size of 32 bits. The pseudocode for the hash follows:

\begin{verbatim}
const int m = 1540483477;

int hash(int[] k, int length)
{
    int h = k ^ length;

    for(int i = 0; i < length; i++) {
        k[i] *= m;
        k[i] ^= k[i] >> 24;
        k[i] *= m;

        h *= m;
        h ^= k[i];
    }

    return h;
}
\end{verbatim}

\subsection{The Jenkins Hash}

Jenkins hash is a family of hash function by Jenkins \cite{jenkins}, but we refer specifically to the ``one-at-a-time'' version. It processes the input in 8-bit blocks, and doesn't use any magic numbers. The pseudocode for the hash follows:

\begin{verbatim}
int hash(int input)
{
    int ret = 0;

    for each byte i in input {
        ret += i;
        ret += (ret << 10);
        ret ^= (ret >> 6);
    }

    ret += (ret << 3);
    ret ^= (ret >> 11);
    ret += (ret << 15);

    return ret;
}
\end{verbatim}

\section{Hash implementations on the GPU}

Each hash can be implemented in a shader for direct evaluation. The only problem is generated by hashes which operate in blocks smaller than 32-bits, because the extension specification only allows 32-bit integers.\\

To overcome this limitation we split the input into 8-bit blocks stored in 32-bit integers, using bit operations, and process those blocks as they were 8-bit integers. This wastes some computational power in the process of splitting and processing bigger integers than it is necessary.

Each hash implementation using GLSL can be seen in Figures \ref{code:glslFNVHash}, \ref{code:glslJenkinsHash} and \ref{code:glslMurmurHash}. All hashes are evaluated for a 32-bit input.

\begin{figure}[t]
\begin{verbatim}
const int prime = 16777619;
const int offset = -2128831035;

int fnv1Hash(int key)
{
    int ret = offset;

    int b0 = (key & 255);
    int b1 = (key & 65280) >> 8;
    int b2 = (key & 16711680) >> 16;
    int b3 = (key & -2130706432) >> 24;

    ret *= prime;
    ret ^= b0;
    
    ret *= prime;
    ret ^= b1;
    
    ret *= prime;
    ret ^= b2;
    
    ret *= prime;
    ret ^= b3;

    return ret;
}
\end{verbatim}
\caption{GLSL implementation of the FNV1 hash}
\label{code:glslFNVHash}
\end{figure}

\begin{figure}[t]
\begin{verbatim}
int jenkinsHash(int key)
{
    int hash = 0;
    
    int b0 = (key & 255);
    int b1 = (key & 65280) >> 8;
    int b2 = (key & 16711680) >> 16;
    int b3 = (key & -16777216) >> 24;
    
    hash += b0;
    hash += (hash << 10);
    hash ^= (hash >> 6);
    
    hash += b1;
    hash += (hash << 10);
    hash ^= (hash >> 6);

    hash += b2;
    hash += (hash << 10);
    hash ^= (hash >> 6);

    hash += b3;
    hash += (hash << 10);
    hash ^= (hash >> 6);
    
    hash += (hash << 3);
    hash ^= (hash >> 11);
    hash += (hash << 15);
    
    return hash;
} 
\end{verbatim}
\caption{GLSL implementation of the Jenkins hash}
\label{code:glslJenkinsHash}
\end{figure}

\begin{figure}[t]
\begin{verbatim}
const int m = 1540483477;

int murmurHash(int k)
{
    int h = 10;

    k *= m;
    k ^= k >> 24;
    k *= m;

    h *= m;
    h ^= k;

    return h;
}
\end{verbatim}
\caption{GLSL implementation of the Murmur hash}
\label{code:glslMurmurHash}
\end{figure}

\section{Partial Hashing}

Initial performance measures using the three chosen hashes showed that the implementation is significantly slower than texture based Perlin Noise. To improve performance, we modified the FNV1 and Jenkins hashes to operate directly in 32-bit integers, instead on 8-bit integers. For the Jenkins hash, we found that sufficient randomness is generated using only one iteration of the inner loop, but for the FNV1 hash, two iterations are required to get smooth  noise. We call this modified hashes ``Partial'' versions. Implementations are shown in Figures \ref{code:glslPartialFNV1Hash} and \ref{code:glslPartialJenkinsHash}.

\begin{figure}[t]
\begin{verbatim}
int hash(int key)
{
    int ret = offset;

    ret *= prime;
    ret ^= key;

    ret *= prime;
    ret ^= key;

    return ret;
}

\end{verbatim}
\caption{GLSL implementation of the Partial FNV1 hash}
\label{code:glslPartialFNV1Hash}
\end{figure}

\begin{figure}[t]
\begin{verbatim}
int hash(int key)
{
    int hash = 0;

    hash += key;
    hash += (hash << 10);
    hash ^= (hash >> 6);

    hash += (hash << 3);
    hash ^= (hash >> 11);
    hash += (hash << 15);

    return hash;
}

\end{verbatim}
\caption{GLSL implementation of the Partial Jenkins hash}
\label{code:glslPartialJenkinsHash}
\end{figure}

\section{Statistical Properties}

Our proposed noise functions generate pseudorandom numbers in $[-1, 1]$, with an average of $0.0$. Perlin noise has a approximate uniform distribution \cite{imageSynthesizer}, but changing the gradient generation might produce a different distribution. We found through simulation that all proposed functions have gaussian-like distributions. Partial hashes produce the same gaussian distribution in the noise output.

Classic Perlin noise has a period of 256 units, which is limited by the size of the lookup tables. Our proposed noise functions don't have a period set by the algorithm, but we choose to limit the period to $2^{20}$ to avoid artifacts because of integer to floating point convertion. 
  
\section{Implementation}

\begin{figure*}[!ht]
  \centering

  \subfloat[Perlin/FNV1]    {\includegraphics[width=0.22\textwidth]{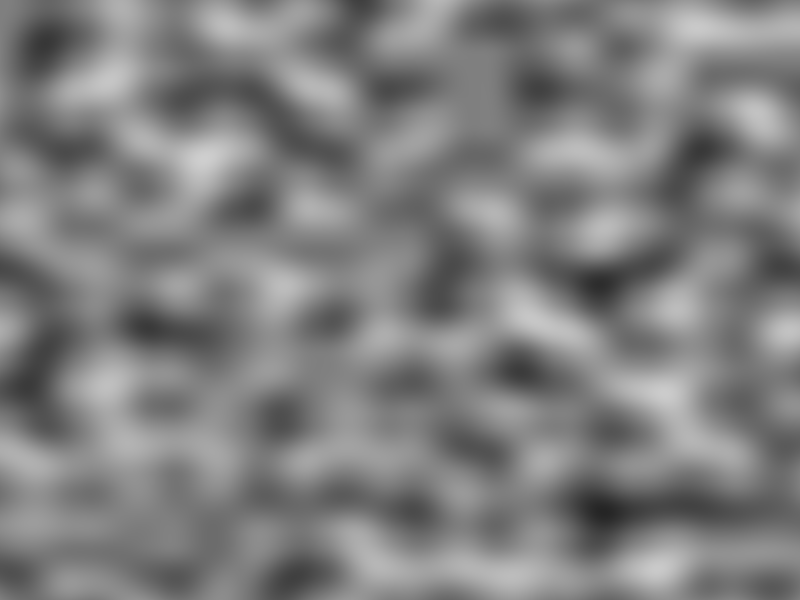}} \,
  \subfloat[Turbulence/FNV1]{\includegraphics[width=0.22\textwidth]{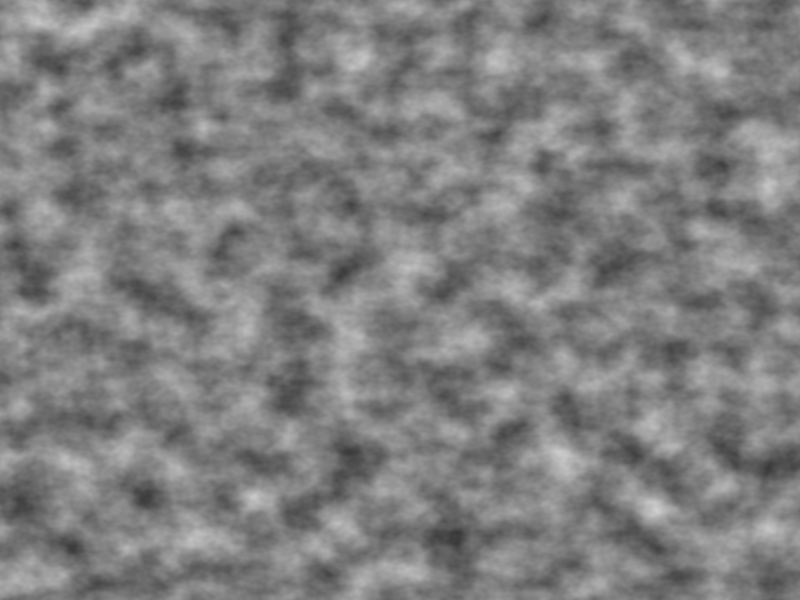}} \,
  \subfloat[Clouds/FNV1]    {\includegraphics[width=0.22\textwidth]{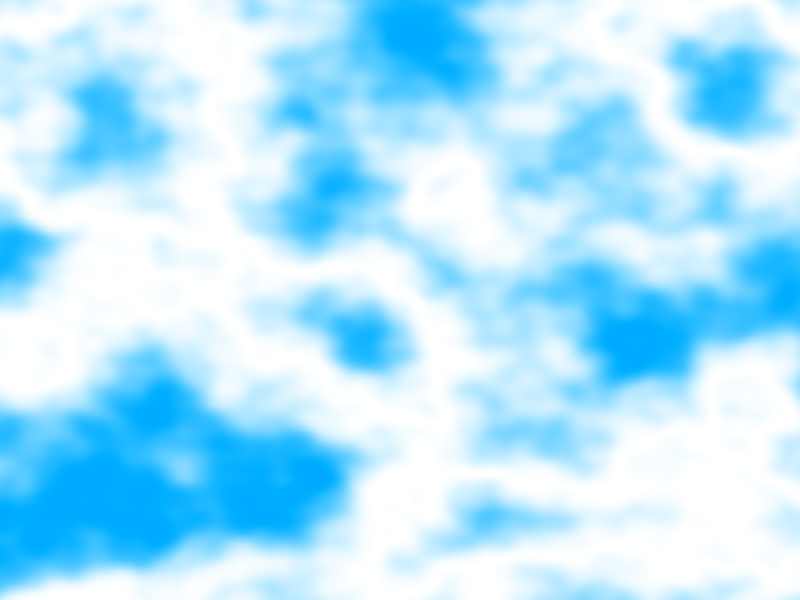}} \\

  \subfloat[Perlin/PartialFNV1]    {\includegraphics[width=0.22\textwidth]{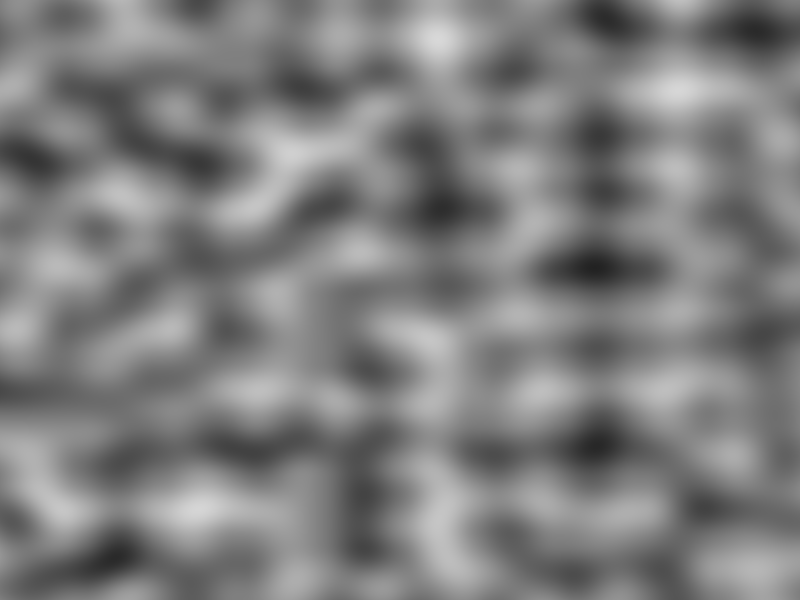}} \,
  \subfloat[Turbulence/PartialFNV1]{\includegraphics[width=0.22\textwidth]{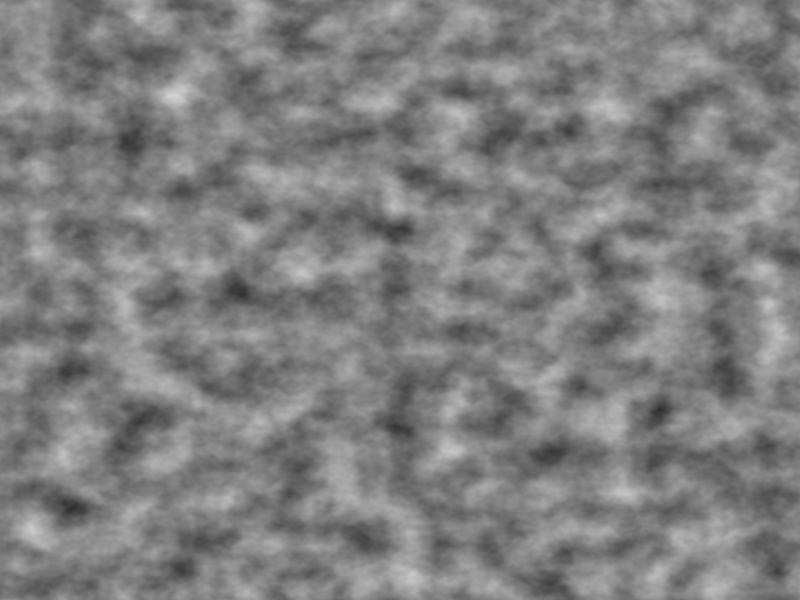}} \,
  \subfloat[Clouds/PartialFNV1]    {\includegraphics[width=0.22\textwidth]{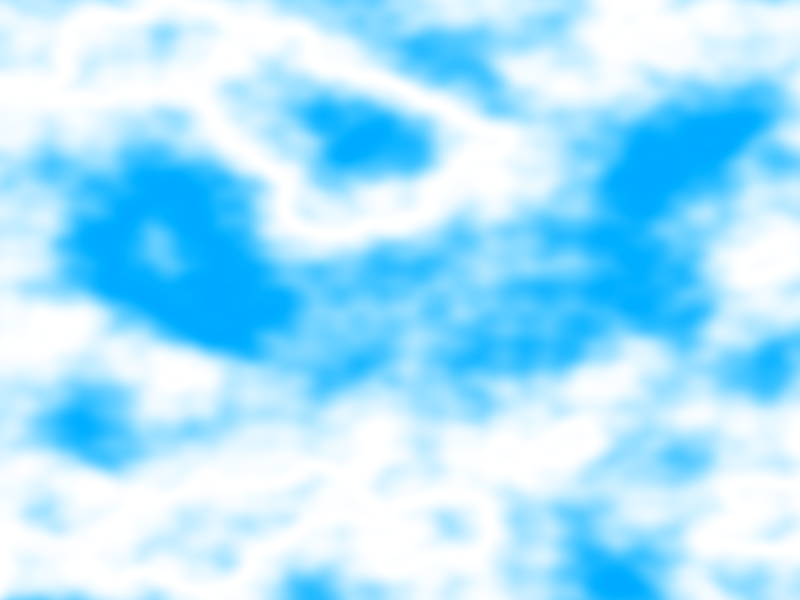}} \\

  \subfloat[Perlin/Jenkins]    {\includegraphics[width=0.22\textwidth]{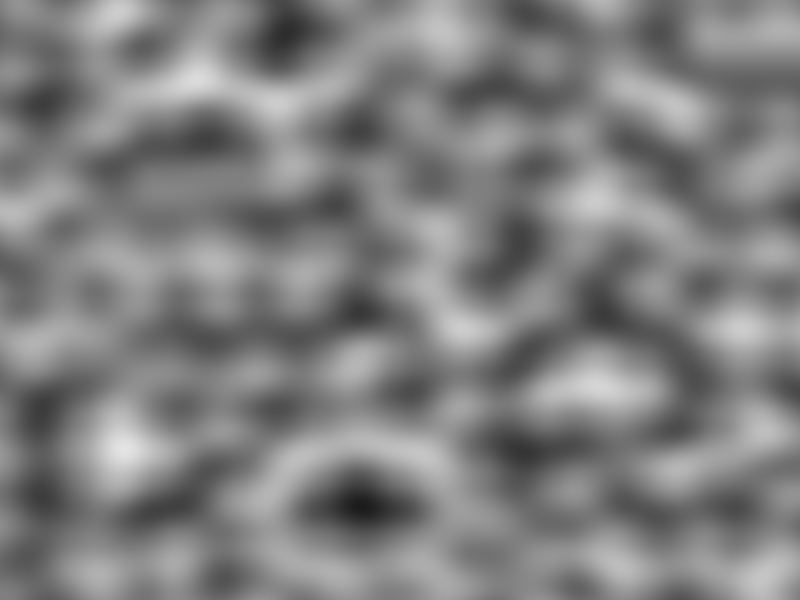}} \,
  \subfloat[Turbulence/Jenkins]{\includegraphics[width=0.22\textwidth]{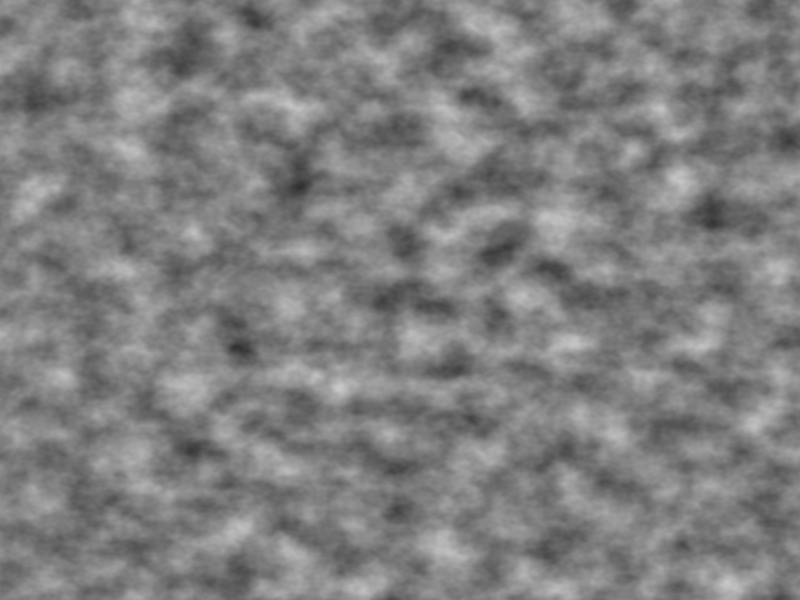}} \,
  \subfloat[Clouds/Jenkins]    {\includegraphics[width=0.22\textwidth]{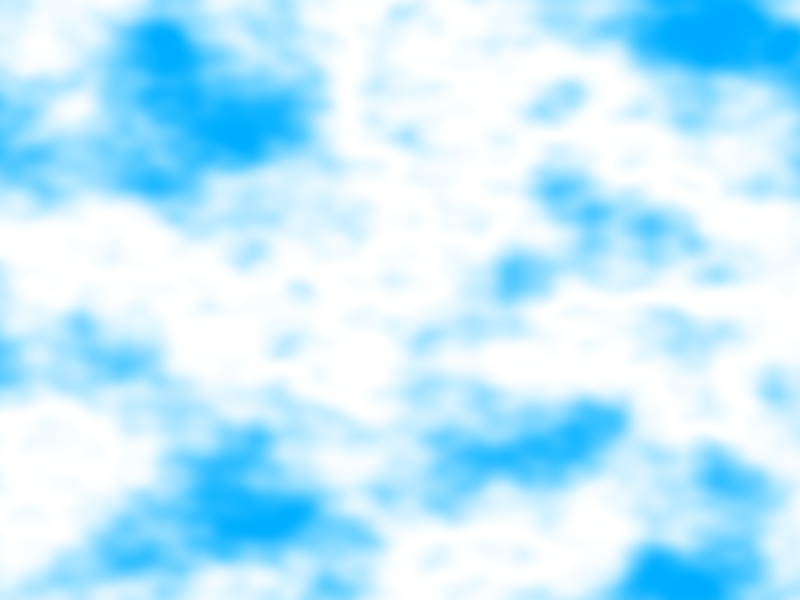}} \\

  \subfloat[Perlin/PartialJenkins]    {\includegraphics[width=0.22\textwidth]{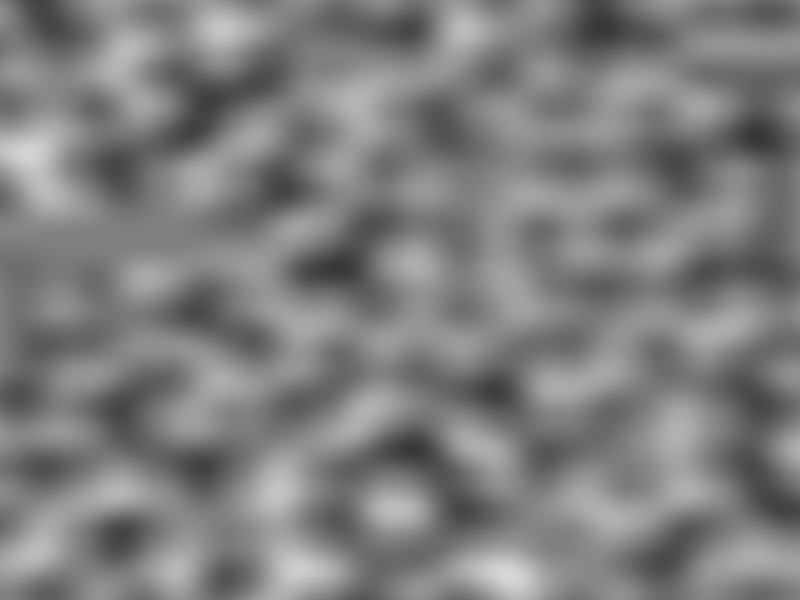}} \,
  \subfloat[Turbulence/PartialJenkins]{\includegraphics[width=0.22\textwidth]{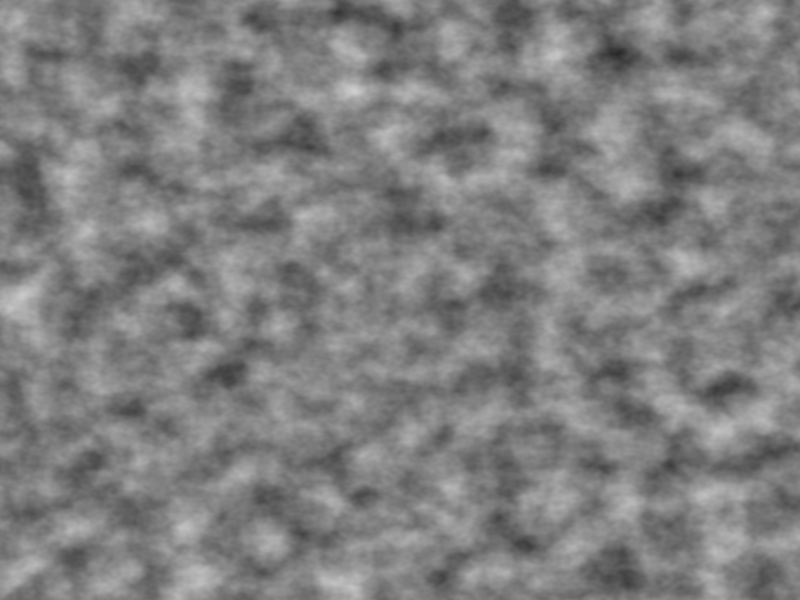}} \,
  \subfloat[Clouds/PartialJenkins]    {\includegraphics[width=0.22\textwidth]{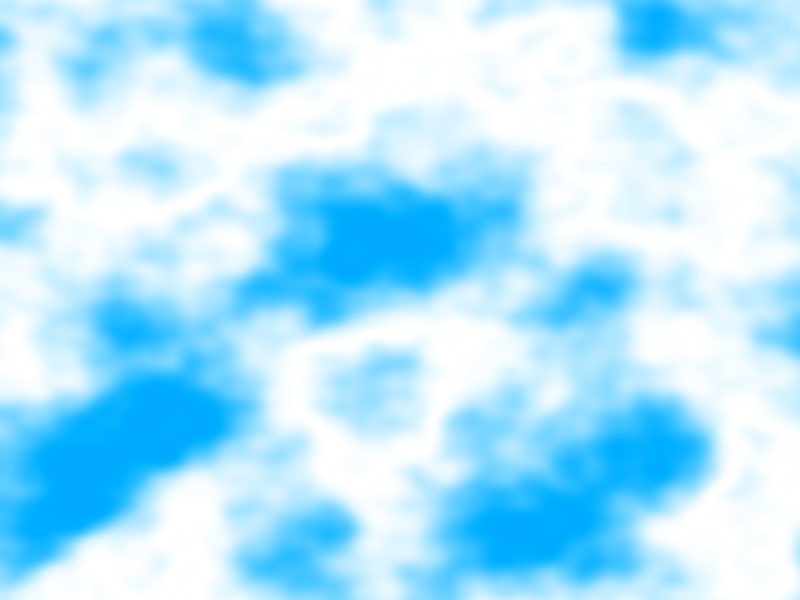}} \\

  \subfloat[Perlin/Murmur]    {\includegraphics[width=0.22\textwidth]{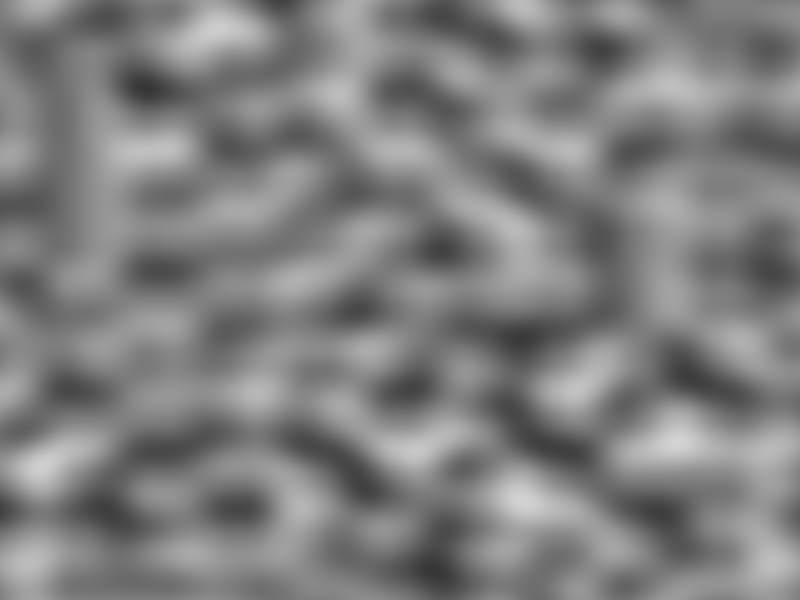}} \,
  \subfloat[Turbulence/Murmur]{\includegraphics[width=0.22\textwidth]{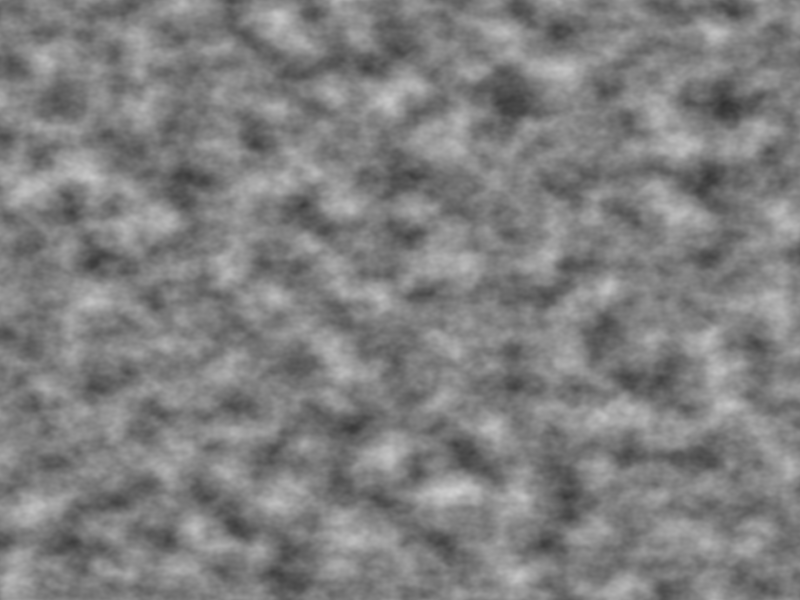}} \,
  \subfloat[Clouds/Murmur]    {\includegraphics[width=0.22\textwidth]{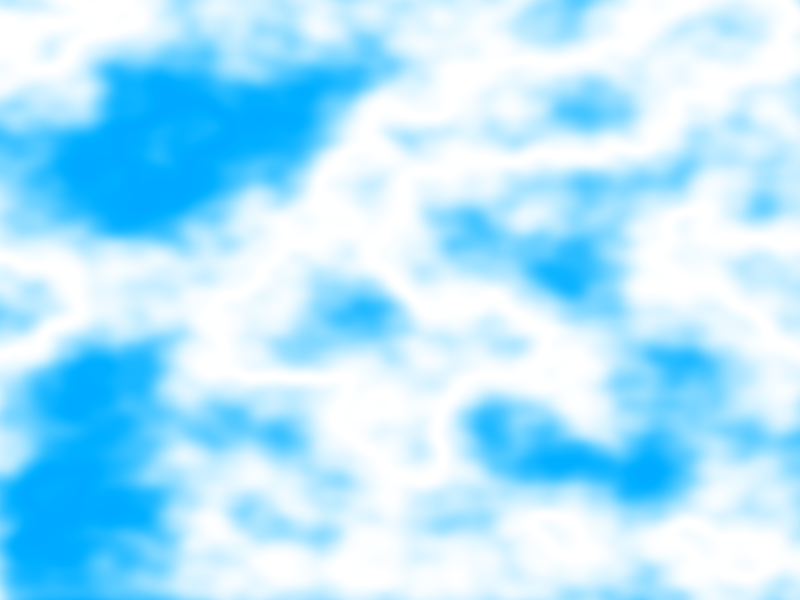}} \\

  \subfloat[Perlin/Float]    {\includegraphics[width=0.22\textwidth]{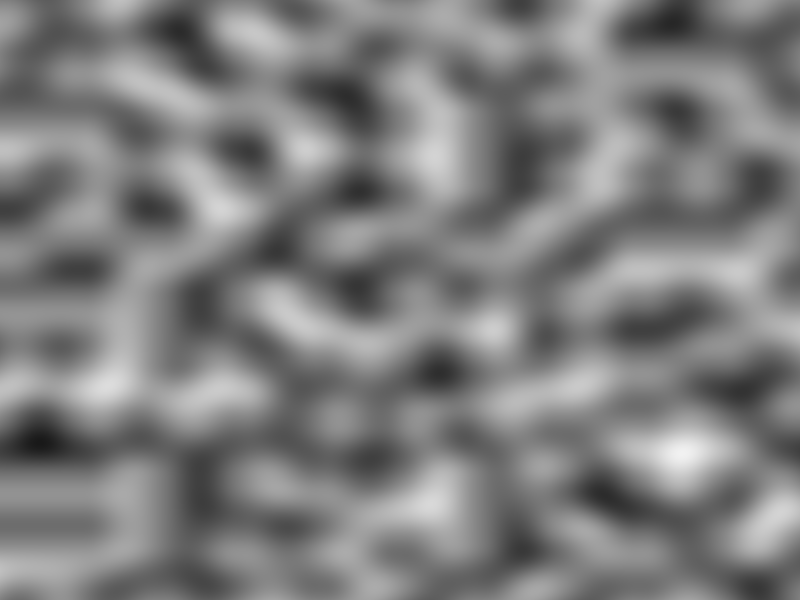}} \,
  \subfloat[Turbulence/Float]{\includegraphics[width=0.22\textwidth]{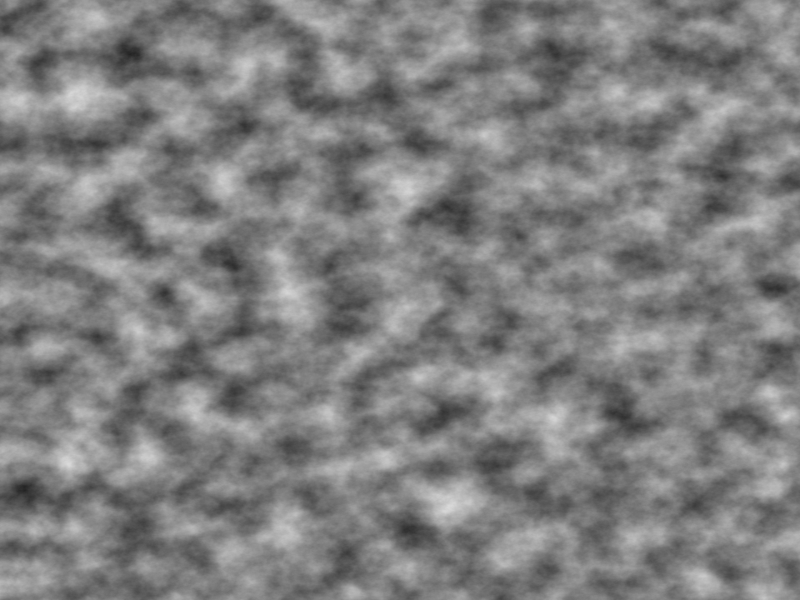}} \,
  \subfloat[Clouds/Float]    {\includegraphics[width=0.22\textwidth]{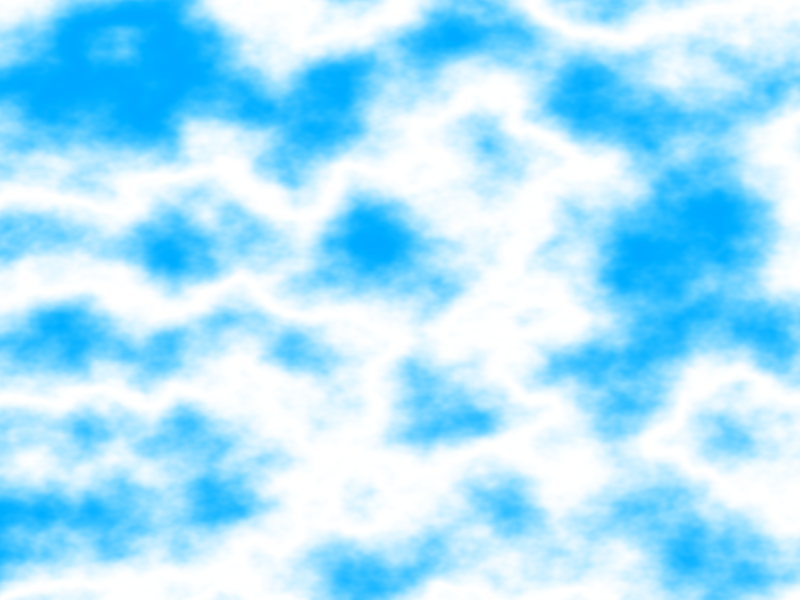}}
  
  \caption{Example renders}
  \label{fig:noises}
\end{figure*}

The proposed modifications were implemented in OpenGL 3.0, using the OpenGL Shading Language v1.30. To measure performance, we rendered a texture mapped quad, using a texture coordinate as input to the noise function; the scalar result was propagated to the rgb components to achieve a grayscale output.
\section{Performance}

Performance measures were made using a Dell XPS m1330 laptop, with a GeForce 8400M GS GPU with 180.37.05 drivers on ArchLinux i686. To get instruction counts, we used NVIDIA's Cg Compiler, which can compile GLSL code to NVfp4 Assembly. 

To get comparable results, an already implemented Perlin Noise function was used. This function is implemented using textures to store the permutation and gradient tables. Two versions of this function were used, one implemented using floating point mathematic (Perlin/Float), and other using integer arithmetic (Perlin/Integer).

Performance was measured using the render time in milliseconds as metric, at different resolutions for 2D and 3D Noise.

\section{Results}

Example renders are shown in Figure \ref{fig:noises}. The first column is a render of the noise function, the second column is a render of the turbulence funciton using the same noise function, and the third column is a render of a proocedural cloud texture using the same noise function. Performance measures are shown in Figure \ref{table:performanceResults}.

Our performance data showed that the proposed implementations are slower than regular texture-based Perlin Noise, only Perlin/PartialFNV1 and Perlin/PartialJenkins are close enough to Perlin/Float to be considered an alternative implementation.

The tradeoff between speed and period is alleviated in Perlin/PartialFNV1 and Perlin/PartialJenkins. Both can be considered alternatives because of their ``cheap'' cost and considerable large period.

Noise generated by our proposed functions is of comparable quality when compared to Perlin/Float and Perlin/Integer.

On great advantage of our implementation is that Modern and newer GPUs can execute more ALU instructions per texture fetches than older processors \cite{atiOpenGLOptimization}, and therefore a developer needs to use more computational power to hide the latency of texture fetching. Our noise functions moves workload from texture bandwith to ALU units, and can help balance the workload between different GPU components.

\section{Future Work}

In the future, we would like to implement other noise functions on GPUs, such as Worley's cellular noise. But more important, is to demonstrate the advantage of hashing functions over precomputed tables in memory bandwith limited applications.

\section{Conclusion}

We researched alternate implementations of noise for modern graphics processing units, using hash functions to replace lookup tables with runtime computable data.

We expect that with faster noise implementations its usage in realtime applications such as commercial games, will grow. We recommend using Perlin/PartialFNV1 and/or Perlin/PartialJenkins as they have a large period ($2^{20}$ units) and its performance is acceptable.

There is still room for improvement. A function can never be fast enough for real time applications. Perlin's Simplex noise could be modified in the same way, but it would only require $O(n)$ contributions from neighbours, as opposed by $2^n$ contributions needed for Perlin noise.

A mix of Simplex noise and hash functions could lead to a silicon hardware implementation. The OpenGL Shading Language specification requires a noise function, but there's no major hardware implementation. The availability of fast noise would push its adoption in the industry.

\begin{figure*}[!htb]
    \centering
    \begin{tabular}{|l|c|c|c|}
        \hline
        2D Noise Algorithm    & Instructions & Rendertime at 800x600 & Rendertime at 1024x768 \\
        \hline
        Perlin/FNV1           & $\sim$ 165  & 11.5 ms & 17.5 ms \\ 
        Perlin/PartialFNV1    & $\sim$ 77   & 6.2 ms  & 8.7 ms  \\
        \hline
        Perlin/Jenkins        & $\sim$ 309  & 14.0 ms & 20.0 ms \\
        Perlin/PartialJenkins & $\sim$ 133  & 7.0 ms  & 10.4 ms \\
        \hline
        Perlin/Murmur         & $\sim$ 93   & 7.5 ms  & 10.5 ms \\
        \hline
        Perlin/Float          & $\sim$ 37   & 4.6 ms  & 8.0 ms  \\
        Perlin/Integer        & $\sim$ 66   & 21.4 ms & 33.1 ms \\
        \hline
        3D Noise Algorithm    & Instructions & Rendertime at 800x600 & Rendertime at 1024x768 \\
        \hline
        Perlin/FNV1           & $\sim$ 473   & 23.6 ms & 33.0 ms \\ 
        Perlin/PartialFNV1    & $\sim$ 209   & 13.3 ms & 19.0 ms  \\
        \hline
        Perlin/Jenkins        & $\sim$ 905   & 29.0 ms & 40.0 ms \\
        Perlin/PartialJenkins & $\sim$ 377   & 15.4 ms & 22.1 ms \\
        \hline
        Perlin/Murmur         & $\sim$ 257   & 17.0 ms & 23.0 ms \\
        \hline
        Perlin/Float          & $\sim$ 77    & 12.5 ms & 17.5 ms  \\
        Perlin/Integer        & $\sim$ 134   & 40.1 ms & 60.0 ms \\
        \hline
    \end{tabular}
    
    \caption{Proposed noise performance measures for 2 and 3 dimensions}
    \label{table:performanceResults}
\end{figure*}

\section*{Acknowledgments}

The authors would like to thank Sebastian Machuca and Gonzalo Gaete.

\bibliographystyle{IEEEtran}
\bibliography{references}

\begin{thebibliography}{10}
\providecommand{\url}[1]{#1}
\csname url@samestyle\endcsname
\providecommand{\newblock}{\relax}
\providecommand{\bibinfo}[2]{#2}
\providecommand{\BIBentrySTDinterwordspacing}{\spaceskip=0pt\relax}
\providecommand{\BIBentryALTinterwordstretchfactor}{4}
\providecommand{\BIBentryALTinterwordspacing}{\spaceskip=\fontdimen2\font plus
\BIBentryALTinterwordstretchfactor\fontdimen3\font minus
  \fontdimen4\font\relax}
\providecommand{\BIBforeignlanguage}[2]{{%
\expandafter\ifx\csname l@#1\endcsname\relax
\typeout{** WARNING: IEEEtran.bst: No hyphenation pattern has been}%
\typeout{** loaded for the language `#1'. Using the pattern for}%
\typeout{** the default language instead.}%
\else
\language=\csname l@#1\endcsname
\fi
#2}}
\providecommand{\BIBdecl}{\relax}
\BIBdecl

\bibitem{imageSynthesizer}
K.~Perlin, ``An image synthesizer,'' \emph{SIGGRAPH Comput. Graph.}, vol.~19,
  no.~3, pp. 287--296, 1985.

\bibitem{glslSpec}
K.~Group, \emph{Open{GL} Shading Language Specification, version 1.40 revision
  5}, 2009.

\bibitem{improvedPerlin}
K.~Perlin, ``Improving noise,'' in \emph{SIGGRAPH '02: Proceedings of the 29th
  annual conference on Computer graphics and interactive techniques}.\hskip 1em
  plus 0.5em minus 0.4em\relax ACM, 2002, pp. 681--682.

\bibitem{olanoModifiedNoise}
M.~Olano, ``Modified noise for evaluation on graphics hardware,'' in \emph{HWWS
  '05: Proceedings of the ACM SIGGRAPH/EUROGRAPHICS conference on Graphics
  hardware}.\hskip 1em plus 0.5em minus 0.4em\relax New York, NY, USA: ACM,
  2005, pp. 105--110.

\bibitem{implementingImprovedPerlin}
S.~Green, ``Implementing improved perlin noise,'' \emph{GPU Gems 2}, 2005.

\bibitem{direct3D10System}
D.~Blythe, ``The {D}irect{3D} 10 system,'' \emph{ACM Trans. Graph.}, vol.~25,
  no.~3, pp. 724--734, 2006.

\bibitem{extGpuShader4}
NVIDIA and Others, ``{GL}\_{EXT}\_gpu\_shader4 {O}pen{GL} extension,'' 2006.

\bibitem{glSpec}
K.~Group, \emph{The OpenGL Graphics System: A Specification, Version 3.0},
  2008.

\bibitem{fnv}
G.~Fowler, L.~C. Noll, and P.~Vo, ``{FNV} hash,''
  \url{http://isthe.com/chongo/tech/comp/fnv/}.

\bibitem{murmur}
A.~Appleby, ``Murmurhash,'' \url{http://murmurhash.googlepages.com/}.

\bibitem{jenkins}
B.~Jenkins, ``A hash function for hash table lookup,''
  \url{http://www.burtleburtle.net/bob/hash/doobs.html}.

\bibitem{atiOpenGLOptimization}
AMD, ``{ATI} {O}pen{GL} {P}rogramming and {O}ptimization {G}uide,'' 2007.

\end{thebibliography}

\end{document}